# The Effect of Structural Phase Changes on Fermi Level Shifts and Optoelectronic Properties of Lead-Free CsSnI$_3$ Perovskites

Dilshod D. Nematov [1, *], Amondulloi S. Burhonzoda [1, **], Mekhrdod S. Kurboniyon [1, 2], Umar Zafari [1, 3], Kholmirzo T. Kholmurodov [1, 6, 7], Tomoyuki Yamamoto [4, 5], Farhod Shokir [1 ***]

[1] S.U. Umarov Physical-Technical Institute of NAST, Dushanbe 734042, Tajikistan
[2] School of Optoelectronic Engineering & CQUPT-BUL Innovation Institute, Chongqing University of Posts and Telecommunications, Chongqing, 400065, China
[3] Center of Innovative Development of Science and New Technologies, NAST, Dushanbe, 734025, Tajikistan
[4] Faculty of Science and Engineering, Waseda University, Tokyo, 169-8555, Japan
[5] Institute of Condensed-Matter Science, Waseda University, Tokyo, 169-8555, Japan
[6] Joint Institute for Nuclear Research, 141980 Dubna, Russia
[7] Dubna State University, 141980 Dubna, Russia
[*] dilnem@mail.ru,  [**] amondullo.burkhonzoda@mail.ru,  [***] farhod0475@gmail.com

**Abstract.** The work carried out first-principles calculations within the framework of density functional theory to study the structural stability of the CsSnI$_3$ compound and the influence of phase transitions on their electronic and optical properties. Using the GGA and SCAN functionals, the relaxed structures of the CsSnI$_3$ phases were obtained and their geometric characteristics were assessed. Using the Phonopy code based on VASP, calculations of phonon and thermodynamic properties were performed, and the temperatures of phase transitions of CsSnI$_3$ were determined. The temperature dependences of the thermodynamic parameters α-, β-, γ- and δ-phases of CsSnI$_3$ were analyzed. The trends in free energy, entropy, enthalpy, heat of formation energy and heat capacity were justified in terms of the pattern of changes in the total energy of the four phases of CsSnI$_3$ from VASP calculations. It is shown that at 0 K the non-perovskite structure of the CsSnI$_3$ compound (δ-CsSnI$_3$) is the most stable (followed by γ-CsSnI$_3$), and the tetragonal phase (β) is quite unstable, having the highest energy among the perovskite phases. It was revealed that at temperatures above 450 K the tetragonal phase becomes stable, and when the temperature drops, it transforms into the cubic phase (α-CsSnI$_3$). The phase transition between the β and γ phase of perovskite occurs in the range of 300-320 K, and at 320 K a black-yellow transformation of CsSnI$_3$ occurs in which the cubic phase (black perovskite) undergoes a phase transition to a non-perovskite conformation (yellow phase). The presence of temperature phase transitions between two orthorhombic phases of CsSnI$_3$ at 360 K was discovered, although direct transitions of the α↔γ and γ↔δ types have not yet been reported in any experiment, except for γ→δ transitions under the influence of moisture. Based on well-relaxed structures (from the SCAN calculations), the band gap widths for four CsSnI$_3$ phases were calculated and compared with experimental measurements. Electronic properties and Fermi level shifts as a result of phase transformations of CsSnI$_3$ were assessed using the HSE06 functional and machine learning prediction. The values of the complex dielectric constant and the refractive index of all phases of the CsSnI$_3$ were determined.

**Keywords**: lead-free perovskites, instability, phase transitions, thermodynamic characteristics, Fermi level shift, electronic and optical properties, density function theory, phonopy calculations, photovoltaic applications.

## 1. Introduction

Over the past few years, metal halide perovskites have intensively attracted the attention of a wide range of researchers and industrial enterprises around the world. They are used in various commercial and technological applications such as solar cells, catalysts, light emitting diodes (LEDs), lasers, X-ray detectors, photodetectors, and field-effect transistors [1-9]. Halide perovskites have also become widely used and recommended in the production of LEDs and luminous bodies with light pumping in the form of luminescent materials [9, 10].

The increasing demand in this area is justified by the fact that perovskite-based semiconductor functional materials have exceptional physical and chemical properties, such as tunable energy band gap, low reflectance, fairly broad absorption spectrum and high absorption coefficient, good photoconductivity and high charge carrier mobility, low exciton binding energy and long diffusion lifetime, optimal electron-hole diffusion lengths and ferroelectricity [6-8, 11-3]. In recent years, they have been widely used as the main raw materials in the absorption layers of solar converters and are actively participating in the program of a universal merciless fight against environmental pollution, reducing the share of carbon dioxide emissions. Along with other materials, perovskites are also actively involved in the program to reduce the rate of use of the earth's depleting fossil fuels in the long term, since the massive burning of fossil fuels in recent years has led to the release of huge amounts of greenhouse gases such as $CO_2$ and $CH_4$ into the atmosphere. To reduce greenhouse gas emissions and ensure energy independence, authorities in major powers are expressing growing interest in developing new alternatives to renewable (clean or low-carbon) energy sources to avoid the catastrophic consequences of global warming in the near future.

Among the several types of clean energy sources available around the world, solar energy is the most promising and promising. According to NREL, in recent years, the efficiency of perovskite solar cells (PCE) has increased quite significantly from 3.8% to more than 26.1% [14]. For layered lead-based perovskites ($MAPbI_3$) solar cells, the highest reported PCE is 25.2% [15]. However, these materials exhibit instability under environmental conditions caused by humidity, humidity, temperature and ultraviolet (UV) light [16]. In addition, Pb-containing perovskites $(MA,Cs)PbX_3$ (X = I, Br, Cl, F) have relatively low dielectric constants, due to which the rate of charge recombination increases and deteriorates the performance characteristics of solar cells, which is the main obstacle to applications of solar panels and cellular devices [17]. Another problem is the presence of lead (Pb), which is toxic and potentially hazardous to the environment [18]. Because of this, it is extremely important to move on to the development and improvement of the properties of lead-free perovskites as an alternative to perovskites containing toxic lead, which supports the EU regulation on the ban and restrictions on the use of compounds containing lead (Pb) in all electronic and electrical devices due to its toxicity impact [18], which corresponds to the goals of the UN sustainable development strategy, in particular SDGs 7 and 13 [19, 20].

In recent years, lead-containing perovskites have been replaced by alternative and lead-free new metal halide organic-inorganic materials, which are being intensively worked on by researchers in the field of materials science. Inorganic perovskites produced by replacing Pb with Ge and Sn have attracted attention due to their better conductivity and absorption than lead-based perovskites. However, it is argued that other problems exist for some of these compounds. For example, it became known that $CsGeI_3$ has brittle behavior, and $CsSnBr_3$ has plasticity [21, 22]. There are a number of other candidates, including yellow-phase compounds based on selenium and tin trihalides ($\delta$-$CsSnI_3$, $\delta$-$CsSnBr_3$, $\delta$-$CsSnCl_3$ and $\delta$-$CsSnF_3$), devoid of the above disadvantages, but characterized by large band gaps, due to which the absorption capacity of the material. In principle, the band gap can be adjusted by changing the composition and doping with foreign ions [23], influencing hydrostatic pressure [24-26] or temperature-induced phase transitions [27,28]. Among these compounds, the most promising material is $CsSnI_3$. Black low-bandgap modifications of $CsSnI_3$ are well suited for photovoltaic devices, but the problem of low stability hinders further progress in this direction. To improve the properties of $CsSnI_3$ and successfully advance in this direction, it is necessary to develop a strategy for the appropriate introduction of external influences, including the influence of temperature and doping, so that along with increasing stability, the band gap can also be controlled and optimized. Regarding the dependence of the perovskite band gap on its composition, in recent years it has been proven that the stability and band gap in any of the metal halide perovskites of the general formula $ABX_3$ strongly depend on the interaction of "B" and "X" of the group X = I, Br, Cl , F) and increase with increasing electronegativity of the "X" cation, which in turn leads to a decrease in the length of the B–X bond [29]. On the other hand, it has been shown that replacing the "A" position with another element does not have a significant effect on the band gap, but through the lattice parameter it mediates the consequences and patterns of the B–X interaction, which sometimes accompanies a doping-induced phase transition [30]. In addition, for a detailed study of doping-induced phase transitions and their influence on the change in the band gap of $CsSnI_3$, it is first necessary to understand the nature of the influence of temperature-dependent phase transitions on the electronic structure and behavior of the Fermi level. In general, the influence of the thermodynamic parameters of perovskites on their electronic and optical properties has become a promising topic of research in recent decades, since many technological applications of these materials and the stable operation characteristics of devices based on them are directly related to the thermal and thermodynamic properties of the raw materials from which these materials are created. devices.

Along with experimental measurements, the properties of perovskites have recently been studied by various theoretical methods, as a result of which the efficiency of solar cells based on them is constantly increasing. One such powerful theoretical approach is Density Functional Theory (DFT), which has become a major tool for the theoretical study of solid materials over the past 10 years, as this powerful approach provides a highly accurate reformulation of quantum mechanical calculations of solids and account for the behavior of electrons in all atomic-molecular environments. This is due to the fact that the Kohn-Sham equations are effectively solved using modern computing clusters [31]. On the other hand, these equations are based on one approximation, namely the exchange-correlation energy, which is responsible for the accuracy of quantum calculations.

In this paper, aspects of structural stability, electronic and optical properties of lead-free perovskite based on $CsSnI_3$ are investigated using DFT calculations. With the help of exchange-correlation functionals SCAN and HSE06 the issues of phase transitions in $CsSnI_3$ and their influence on optoelectronic properties of $CsSnI_3$ are studied, for detailed understanding of the nature of their electronic structure and expedient choice of doping element stabilizing $CsSnI_3$ under environmental conditions as a promising candidate for solar cells with increased efficiency.

## 2. Computational Details

The structural, electronic and optical properties of the α-, β-, γ- and δ-phases of $CsSnI_3$, were investigated based on density functional theory (DFT). Calculations were carried out in the VASP plane wave package [32]. The crystal structures of α-$CsSnI_3$ (cubic), β-$CsSnI_3$ (tetrogonal), γ-$CsSnI_3$ (orthorombic) and δ-$CsSnI_3$ (orthorhombic-non perovskite) were first fully optimized taking into account the relaxation of lattice parameters and atomic positions. All four modifications of $CsSnI_3$ were relaxed using the GGA (PBE) [33] and strictly constrained normalized potential (SCAN) functionals [34]. The electronic states $Cs[5s^2 5p^6 6s^1]$, $I[5s^2 5p^5]$ and $Sn[5s^2 5p^2]$ were considered as valence electrons. After performing a series of convergence tests, the kinetic energy cutoff value was set to 450 eV for all four $CsSnI_3$ phases, and 8×8×8, 6×6×8, 5×5×4, and 5×10×3 k-points according to the Monkhrost-Pack scheme were chosen for geometric optimization of the α-, β-, γ- and δ- phases of $CsSnI_3$. However, the cutoff energy value for calculations of electronic, thermodynamic and optical properties was increased to 800 eV. Calculations of phonon dispersion and thermodynamic properties were performed using the Phonopy [35] code at smaller k-point values, since such calculations are computationally expensive, especially for larger systems with low symmetry. The forces were estimated on supercells 2×2×2 (α), 2×1×2 (β), 1×1×2 (γ), 1×2×1 (δ), for which VASP was used as a calculator. Phonopy calculations are performed on a 40-atom supercell using reduced k-point grids (6×6×6, 6×5×5, 3×2×3, and 2×4×2 for α, β, γ, and δ- phases of the $CsSnI_3$ compound, respectively), and the phonon frequencies were estimated selected on an interpolated grid of 32×32×32 q-points (for the α and β phases) and 24×24×24 for the two orthorhombic (γ and δ) phases. The temperatures of phase transitions in the $CsSnI_3$ system were determined by subtracting the calculated free energy of the $CsSnI_3$ phases (DF) transforming into each other, for which DF was taken into account as the sum of Helmhotz free energies from Phonopy calculations with the minimum energy found from VASP calculations.

The band gap values of $CsSnI_3$ were calculated and compared using the exchange-correlation functionals GGA, SCAN and HSE06 [36], however, for a detailed analysis of the electronic structures and optical spectra of $CsSnI_3$, the hybrid functional HSE06 was used, since this promising functional has proven itself well in recent years and has a leading position among other functionalities for characterizing the electronic properties of materials [37]. The Fermi level shift was estimated by determining the difference in the energy of the most accurate electron in the valence band for each phase, at which the maximum of the valence band of γ-$CsSnI_3$ was taken as the reference point.

## 3. Results and discussions

The relaxed geometric characteristics and crystal lattice constants of the α-, β-, γ- and δ-phases of $CsSnI_3$ (with GGA and SCAN calculations) are given in Table 1 and compared with the results of experimental measurements.

Table 1. Relaxed lattice parameters of α-, β-, γ- and δ- modifications of $CsSnI_3$; The calculated results are compared with the experimental results.

| Phase | | α (Cubic) a=b=c (Å) | β (Tetragonal) a=b, c (Å) | γ (Orthorhombic) a, b, c (Å) | δ (Orth. non-perovskite) a, b, c (Å) |
|---|---|---|---|---|---|
| **Lattice constants** | GGA | 6.261 | 8.789, 6.318 | 8.957, 8.667, 12.503 | 10.621, 4.790, 18.893 |
| | SCAN | 6.200 | 8.680, 6.239 | 8.847, 8.533, 12.372 | 10.543, 4.750, 17.882 |
| | EXP.[38] | 6.206 [38] | 8.712, 6.191 | 8.688, 8.643, 12.378 | 10.349, 4.763, 17.684 |
| **Space group** | | Pm3m | P4/mbm | Pnam | Pnma |
| **Sn-I Bond Distances, Å (SCAN)** | Sn-I1 | 3.089 | 3.133 | 3.122 | 3.224 |
| | Sn-I2 | | 3.123 | 3.153 | 3.219 |
| | Sn-I3 | | | | 2.971 |
| **Bond Angles** | | | | | |
| Sn-I1-Sn | | 180° | - | 172.3° | - |
| Sn-I2-Sn | | 180° | 167.99° | 158.1° | - |

According to Table 1, the SCAN functionality describes the geometry much better than the standard GGA-PBE. The results within this potential are in good agreement with experiment, which speaks to the effectiveness of the use of SCAN for the relaxation of such solid-state systems. An example of this can be observed from a comparison of the calculated X-ray diffraction patterns we obtained for γ-CsSnI₃ with the results of Yuanyuan Zhou and others [39], from which it can be seen that the results we obtained are similar to the results of experimental measurements (Fig. 1). According to calculations, interatomic distances (especially Sn-I) change significantly depending on the phase formation of CsSnI₃. Along with this, the Sn-I1-Sn and Sn-I2-Sn bond angles also change.

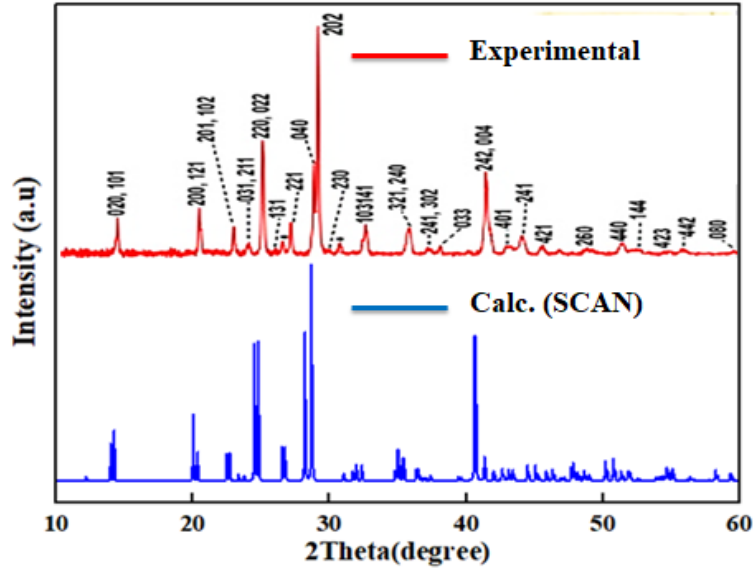

Figure 1. Comparison of theoretical X-ray patterns of γ-CsSnI₃ with X-ray patterns of the orthorhombic phase of CsSnI₃ perovskite [39] obtained by the Bridgman method

Table 2 compares the values of the total energies of the α-, β-, γ- and δ-phases of CsSnI₃, calculated using the GGA functional, from which it is clear that the most stable conformation for CsSnI3 is the δ-modification of this compound. However, γ-CsSnI₃ is the most stable among iodides with a perovskite structure.

Table 2. GGA-calculated total energies of CsSnI₃ phases.

| System | Energy/atom | ΔE |
|---|---|---|
| α-CsSnI₃ | -2,8198 | 0.0108 |
| β-CsSnI₃ | -2,8171 | 0.0235 |
| γ-CsSnI₃ | -2,8201 | 0.0105 |
| δ-CsSnI₃ | -2,8306 | 0 |

According to the results given in Table 2, the CsSnI₃ compound stabilizes as it transitions from the cubic phase to the orthorhombic phase or transitions to a non-porovskite stable structure. Similarly, the heats of formation (ΔH$_f$) for the α-, β-, γ- and δ-phases of CsSnI₃ relative to the constituent elements in their standard states were also calculated in order to evaluate the relative enthalpy stability of CsSnI₃ (Table 3). To do this, we carried out additional calculations for the structures of the cell I (Cmca), Sn (I41/amd) and Cs (Im3m). In our case, we calculated the value of ΔHf in relation to the constituent elements using the following equation:

$$\Delta H_f = E_{tot}^{CsSnI_3} - (E_{tot}^{Cs} + E_{tot}^{Sn} + 3E_{tot}^{I}) \qquad (1)$$

where $E_{tot}^{CsSnI_3}$, $E_{tot}^{Cs}$, $E_{tot}^{Sn}$ and $E_{tot}^{I}$ are the total energy values of the CsSnI₃ phase and the total energies of the pure components Cs, Sn and I in their respective states.

Table 3. Heat of formation of α-, β-, γ- and δ-phases of CsSnI₃, calculated from equilibrium structures at 0 K.

| System | ΔH$_f$/atom |
|---|---|
| α-CsSnI3 | -0,67406 |
| β-CsSnI3 | -0,67135 |
| γ-CsSnI3 | -0,67435 |
| δ-CsSnI3 | -0,68485 |

According to the results obtained, as we move from the cubic to the orthorhombic phase, the heat of formation energy decreases, which indicates the relative stability of the orthorhombic phase of the perovskite in environmental conditions. However, the still non-perovskite structure of CsSnI₃ continues to show a trend of maximum stability among all phases of this compound. Higher values of the heat of formation (negative values) obtained for the γ- and δ-phase of CsSnI₃ indicate a greater amount of energy that is released for the formation of these phases, which ultimately makes them more stable. To confirm the reliability of the results obtained in Tables 1 and 2, the temperature-dependent values of the Helmholtz (or Gibbs) free energy, as well as the relative entropy of the systems under study as a function of temperature, can be obtained after calculating the phonon frequencies and lattice energy of the equilibrium structure [40].

In Figure 2, the graphs of the temperature dependence of the Helmhols free energy (F) for the α-, β-, γ-phases relative to the δ-phase of CsSnI₃, which claims to be the structure with the lowest energy value at 0 K. Also in Figure 3 (a, b ), demonstrates the temperature dependence curves of entropy (ΔS) and heat capacity (C$_v$) for four phases of CsSnI₃.

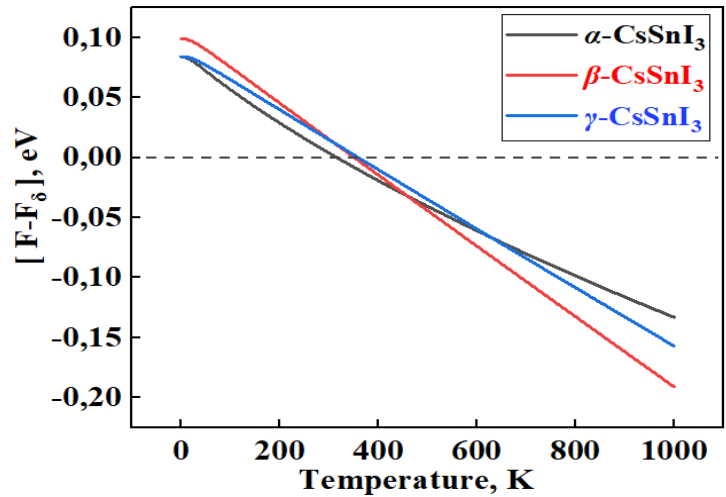

Figure 2. Temperature-dependent difference in Helmholtz free energy for the α-, β-, γ-phase of CsSnI₃ relative to its δ-phase

Figure 4 compares and shows the enthalpy dependence curves (ΔH) of the systems under study, obtained based on the expression:

$$\Delta H = F + \Delta S * T,$$

where T is the absolute temperature. From the results, it can be seen that the β phase has the highest enthalpy and the δ phase is naturally the lowest in all temperature ranges.

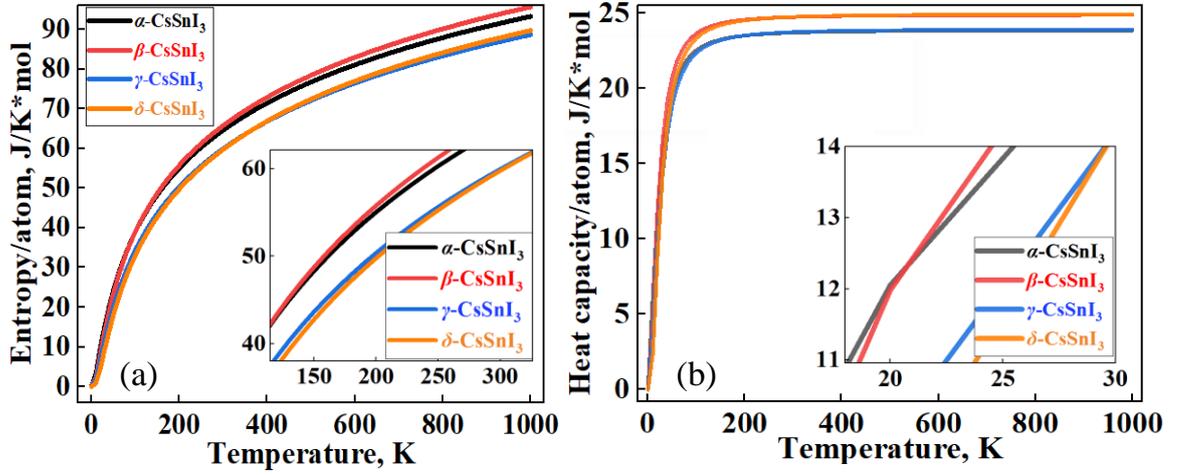

Figure 3. Variation of the entropy (a) and heat capacity (b) depending on temperature for groups of α-, β-, γ- and δ-phases of $CsSnI_3$

According to the results of Figure 2, at 0 K, the β phase has the highest energy among the perovskite phases, and the energy value for the γ phase is the lowest. In this case, the cubic phase is fixed between these phases. However, according to the graph, it is the tetragonal phase that becomes the most stable at high temperatures. The orthorhombic perovskite phase remains energetically close to the α phase up to high temperatures. Calculations show that there is energy competition between the non-perovskite phase of $CsSnI_3$ with the β- and γ-phases of $CsSnI_3$ in the region of 320-360 K. According to the results in Figure 3 (a, b), the free energy of high-temperature phases in almost all temperature ranges is higher than the free energy δ-$CsSnI_3$, which contradicts the rule of direct dependence of free energy and the stability of materials. The general heat capacity picture also shows a similar trend.

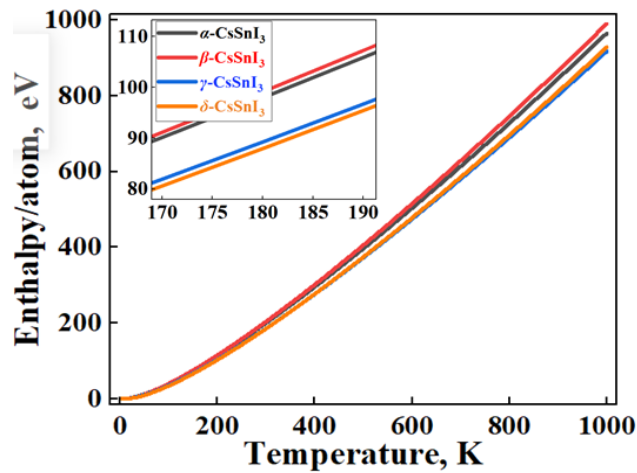

Figure 4. Temperature-dependent enthalpy vibration of α-, β-, γ- and δ-phases of $CsSnI_3$

Next, by subtracting the calculated free energy of the phases ([α-β], [β-γ] and [α-δ]), the phase transition temperatures for $CsSnI_3$ were found. In this case, the free energy was taken into account as the sum of the Helmholtz free energies from Phonopy calculations with the found minimum energy from VASP calculations. Since CsSnI$_3$ phases undergo complex phase transitions at different temperatures, phase transition diagrams were drawn in separate coordinate systems and were also additionally combined into one figure (Figure 5) in order to compare phase transition diagrams in the same range of energy differences.

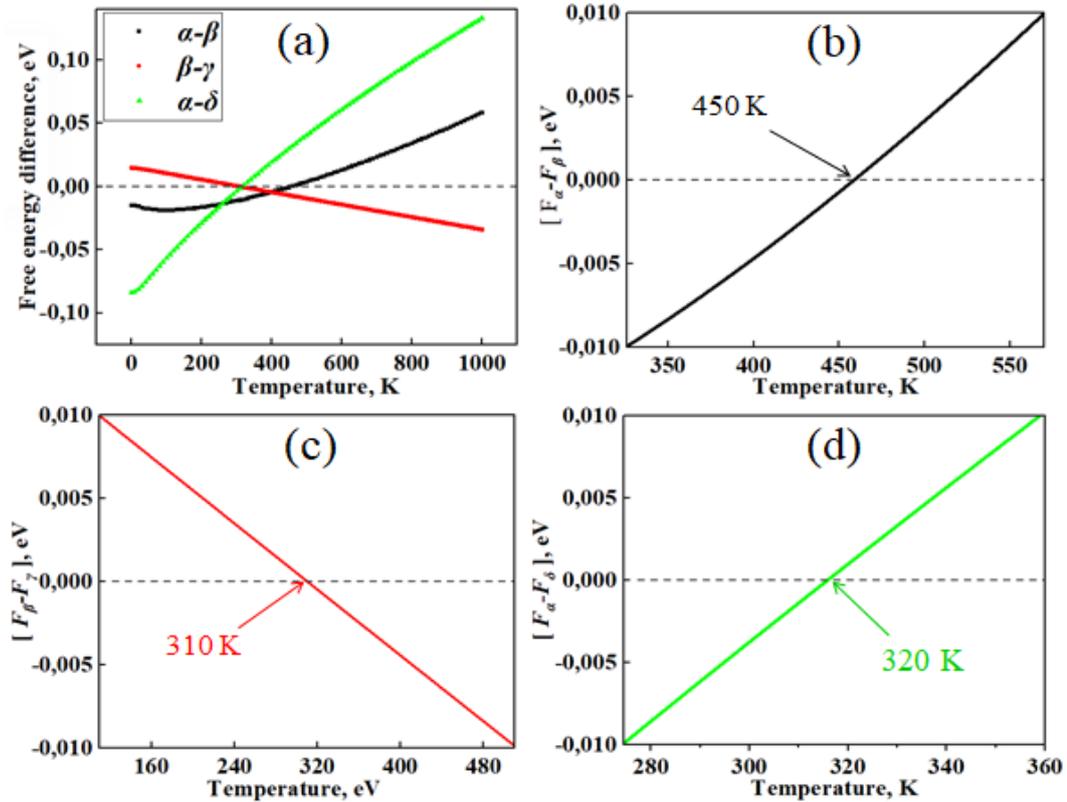

Figure 5. Temperature dependence of the free energy difference at α⇌β, β⇌γ and α⇌δ phase transitions CsSnI$_3$

According to the results shown in Figure 5, CsSnI$_3$ crystals are characterized by three cases of phase transitions in certain temperature ranges. The critical temperature points of phase transitions α⇌β, β⇌γ and α⇌δ indicate that the range of stable existence of these phases differs significantly from each other (Figure 5a). At temperatures above 450 K, the tetragonal phase becomes stable and below this temperature transforms into a cubic conformation (Figure 5b). The phase transition between tetragonal and orthorhombic perovskite occurs in the range of 300-320 K (Figure 5c), and at 320 K a transformation occurs between the CsSnI$_3$ perovskite structure and its non-perovskite analogue, the so-called black-yellow transformation (Figure 5e). The obtained results are similar to the experimental measurements by Koji Yamada and others [41] with the exception of the temperature of phase transitions between the α and δ phases of the perovskite. Calculations also showed the presence of temperature phase transitions between two orthorhombic phases of CsSnI$_3$ at 360 K, although direct transitions of the α ↔ γ and γ ↔ δ types have not yet been detected in any experiment, except for γ → δ transitions under the influence of moisture [42].

First-principles calculations of phonon dispersion in Figure 6 (a-d) show the absence of negative frequencies for the non-perovskite phase of CsSnI$_3$, which indicates its relative stability. This is followed by the orthorhombic perovskite phase.

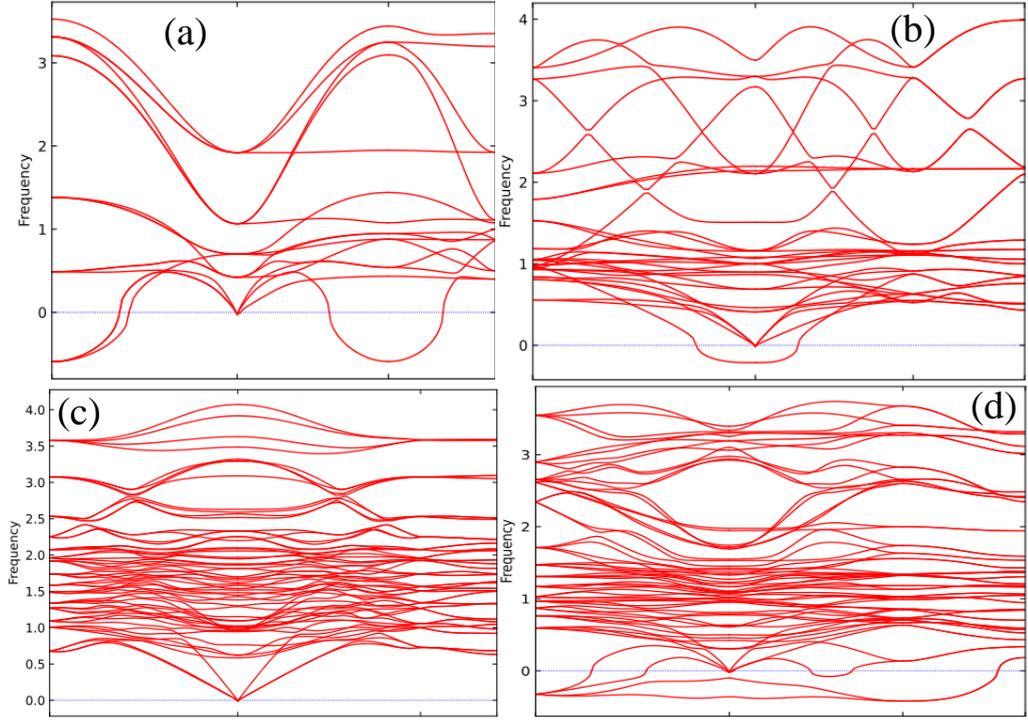

Figure 6. Phonon dispersions for: α-CsSnI$_3$ (a), β-CsSnI$_3$ (b), δ-CsSnI$_3$ (c) and γ-CsSnI$_3$ (d)

Calculations of the density of phonon states also confirm the absence of states in the negative energy region for the non-perovskite phase of CsSnI$_3$ (Figure 7a). Also noticeable is the moderate contribution of the phonon state for the tetragonal phase (Figure 7b), in contrast to the more noticeable state for the cubic (Figure 7a) and tetragonal phase (Figure 7c).

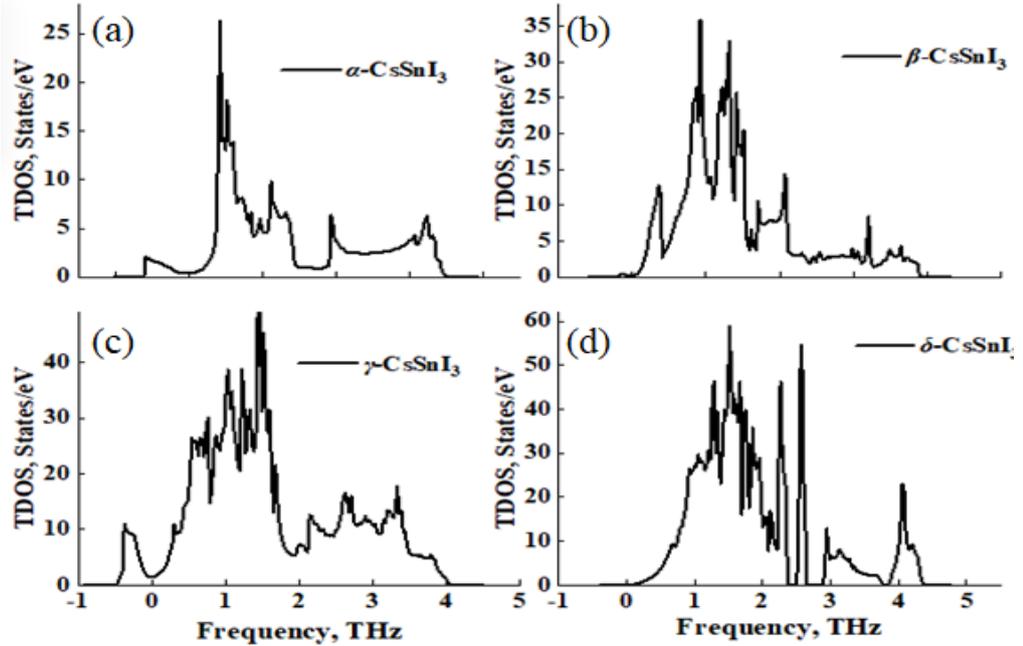

Figure 7. Phonon state densities for: α-CsSnI$_3$ (a), β-CsSnI$_3$ (b), γ-CsSnI$_3$ (c) and δ-CsSnI$_3$ (d)

Next, we studied the electronic and optical properties of the systems under study in order to assess the relationship between the thermodynamic and optoelectronic properties of CsSnI$_3$, as well as demonstrate the influence of stability and phase transitions on the electronic properties, band state, Fermi level shift, absorption and reflection abilities of these compounds, since the assessment The influence of temperature factors on the overall characteristics of the system is critical from the point of view of the use of the material in any device [43, 44].

Thus, using the well-optimized structures of the α-, β-, γ- and δ-phase of CsSnI$_3$, we performed spin-polarized calculations of the band gap of these compounds using the GGA, SCAN and hybrid functional HSE06 (Table 4). The calculated values of the band gap obtained by three different functionals are compared with each other, as well as with the results of machine learning prediction (linear regression method) and data from experiments.

Table 4. Calculated and experimental band gap of α-, β-, γ- and δ- phases CsSnI$_3$ in eV.

| System | This work | | | | Experiment |
|---|---|---|---|---|---|
| | GGA | SCAN | HSE06 | ML Prediction | |
| α-CsSnI$_3$ | 1.116 | 1.122 | 1.326 | 1.232 | - |
| β-CsSnI$_3$ | 0.942 | 1.108 | 1.230 | 0.876 | - |
| γ-CsSnI$_3$ | 1.161 | 1.146 | 1.435 | 1.171 | 1.31 [45] |
| δ-CsSnI$_3$ | 2.178 | 2.488 | 2.990 | 2.753 | - |

According to Table 4, different potentials estimate the band gap differently. In particular, SCAN and GGA (PBE) showed a fairly small band gap compared to HSE06, which is an expensive approach giving results comparable to experiments. However, underestimating the band gap is only a typical error in calculations using these functionals [46]. Moreover, despite the lengthy costs in HSE06 calculations, its continued use is advisable, since the solid-state computing community pays great attention to the problem of correctly predicting the fundamental band gap, since the width of the band gap is one of the main fundamental electrical characteristics of materials. Figure 8 (a-b) shows the dependence of the bandgap width of CsSnI$_3$ on the phase of existence.

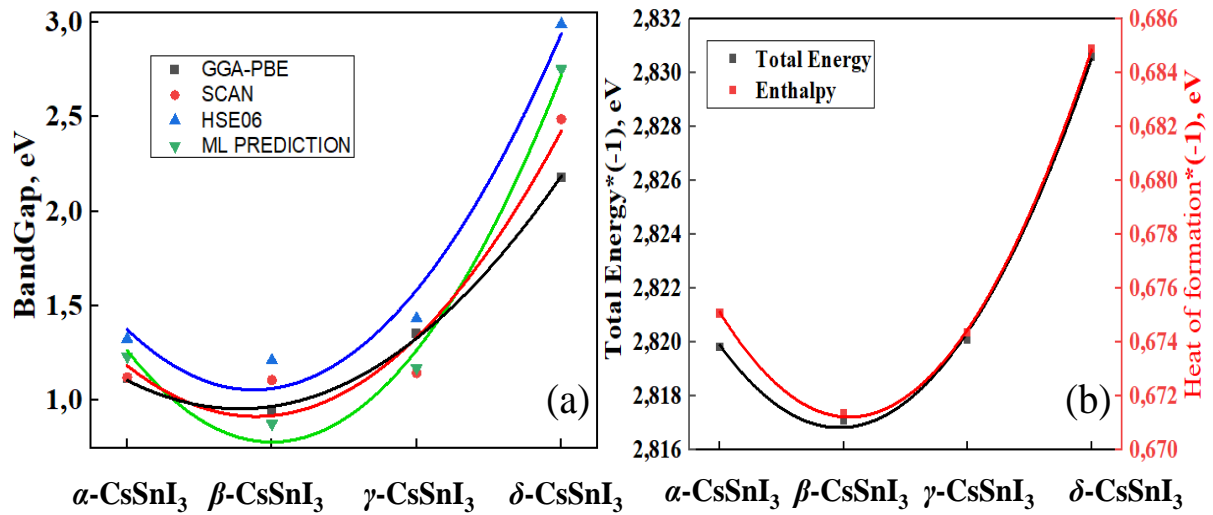

Figure 8. Curves of the band gap (a) and heat of formation (b) depending on the phase formation of CsSnI$_3$

According to the results obtained, phase transformations significantly affect the electronic properties of CsSnI$_3$. As can be seen from Figure 8a, the band gap width decreases as a consequence of the α→β transition, and then begins to increase during the next phase transition in the β→γ region. Similar trends in changes in the width of the forbidden band, the reason for which is a change in the volume of CsSnI$_3$ under the influence of their phase transformations, can be observed from the calculations of GGA, SCAN, HSE06, as well as machine learning predictions for determining the width of the band gap (Figure 8a). The patterns of changes in the band gap are in good agreement with the patterns of changes in the total energy of the $CsSnI_3$ phase and their heat of formation, given in Figure 8b and Tables 2-3. This clearly indicates a special relationship between the band gap width and the thermal properties of the material and their special influence on the electronic structure [47-49].

For a more detailed understanding of these phenomena, we assessed the Fermi level shift as a consequence of the phase transitions of $CsSnI_3$ (Figure 9). These shifts are assessed by determining the change in the energy position of the highest electrons in the valence band for each phase. In this case, the maximum of the valence band of $\gamma$-$CsSnI_3$ was taken as the starting point for comparative analysis.

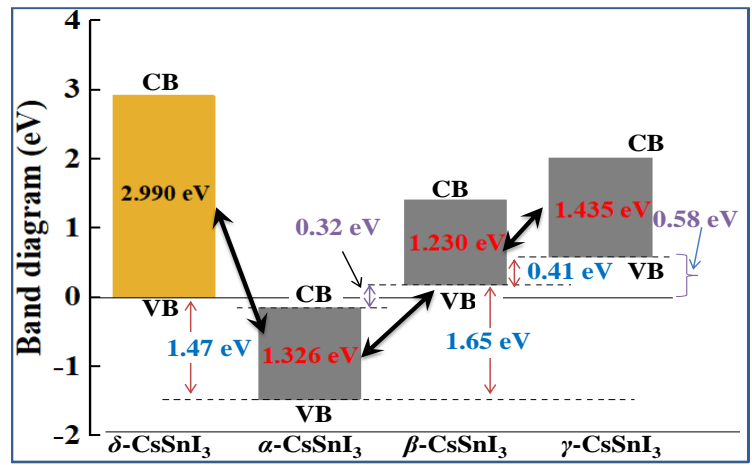

Figure 9. Schematic diagram of the band gap and Fermi level shift during phase transitions of $CsSnI_3$

According to Figure 9, when yellow $CsSnI_3$ is heated and transformed into the cubic perovskite phase, the Fermi level falls towards the low-energy energy range (towards the valence band) and the band gap decreases from 2.99 to 1.326 eV. During the α↔β transition, the band gap once again decreases to 1.23 eV and the Fermi level mixes by 1.65 eV towards the conduction band (CB), and β↔γ transitions increase the band gap to 1.435 eV, and once again times shifts the Fermi level by 0.41 eV in the direction of the conduction band.

Figure 10 (a) compares the total densities of electronic states of the α-, β-, γ- and δ- phases of $CsSnI_3$, from which it can be seen that α-$CsSnI_3$ is characterized by moderate densities of state in the vicinity of the valence band. However, as $CsSnI_3$ moves to lower temperature phases, the density of state increases at the threshold where the valence band (VB) and conduction bands (CB) meet. The PDOS diagram shown in Figure 10b shows that the formation of the conduction band of $CsSnI_3$ is mainly contributed by the s- and p-state of electrons. Also noticeable is a moderate clade of d-orbitals in all phases. It can be seen that the contribution of p-electrons increases sharply with the transition from the perovskite conformation (α-phase) to its non-perovskite analogue (δ).

Figure 11 (a-e) summarizes the results of calculations of the optical properties of α-, β-, γ- and δ- phases of $CsSnI_3$, from which it can be seen that the behavior of the spectra of optical absorption (α), photoconductivity (σ) and energy loss function (ELF) obey the trend of changes in the band gap due to volume vibration as a result of $CsSnI_3$ transitions.

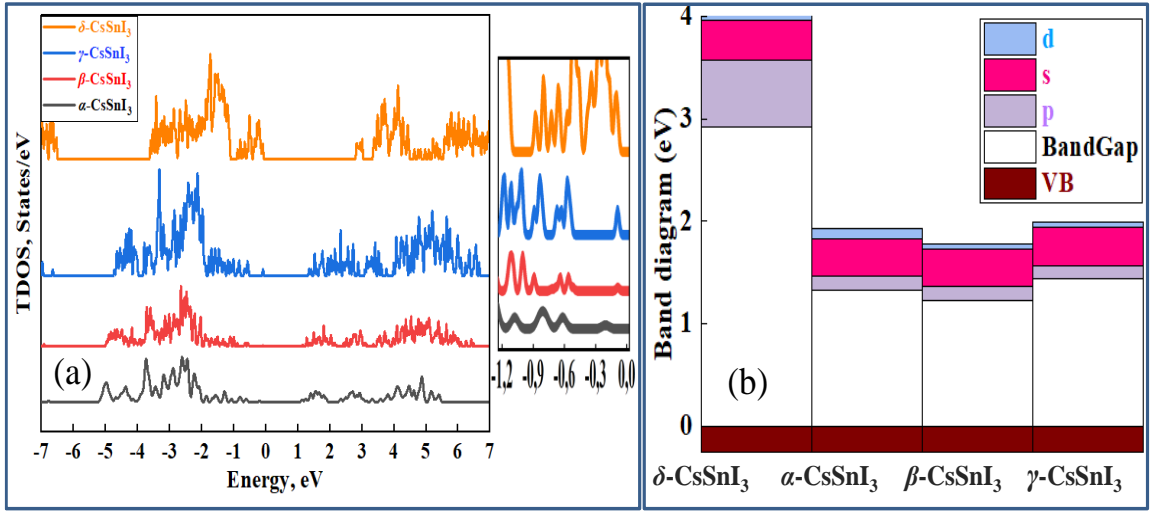

Figure 10. Comparison of (a) total state densities and (b) diagram of the contributions of the s-, p-, and d- orbitals of the α-, β-, γ-, and δ- phases of $CsSnI_3$ in the conduction band. The top of the VB is scaled to zero

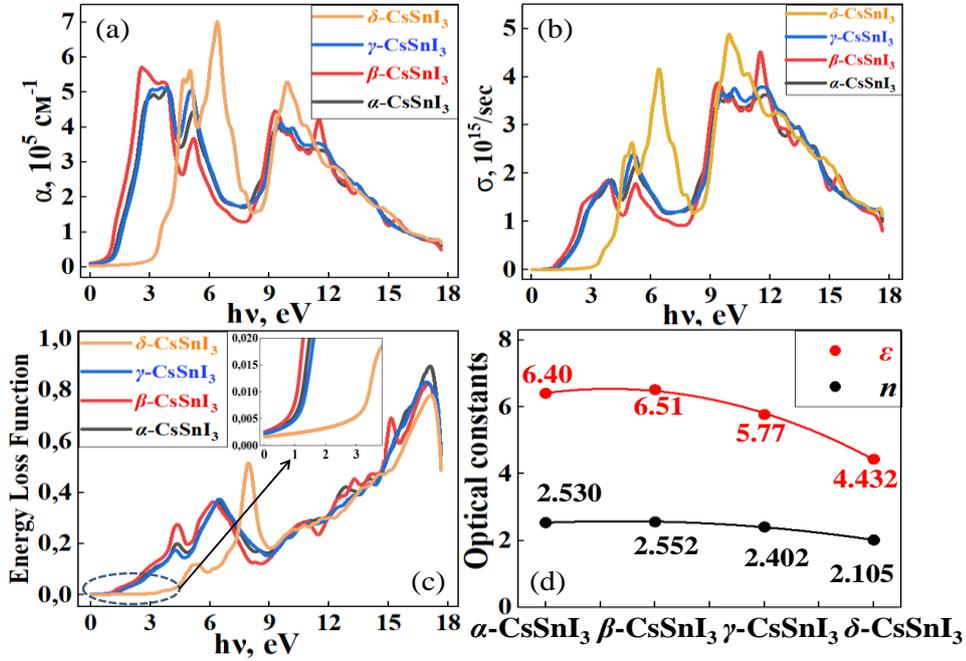

Figure 11. Curves of the absorption coefficient (a), optical conductivity (b) and energy loss function (c) depending on the photon energy for α-, β-, γ- and δ- phases of $CsSnI_3$. Figure (e) summarizes the optical constants of static permittivity (ε) and refractive index (n) depending on the phase formation of $CsSnI_3$.

From the spectra shown in Figure 11 (a-c), it is clear that as we move from the high-temperature phase to the lower-temperature phase, absorption and photoconductivity deteriorate in the infrared (IR) range of light wavelengths, and when moving to non-perovskite conformation, the $CsSnI_3$ compound can absorb only short-wavelengths (high energy) rays of light. High absorption and optical conductivity indicate that all $CsSnI_3$ crystals with a perovskite structure have ideal spectral characteristics that are well suited for photovoltaic applications, however, moving to a more stable phase, $CsSnI_3$ is activated only in the ultraviolet range of electromagnetic waves. However, by taking any necessary measures to stabilize $CsSnI_3$ without significant changes in the bandgap and loss of their absorption capacity, it is possible to obtain unique materials as the main absorption words in new generation solar panels. Figure (e) shows the phase dependence of optical constants using the example of static permittivity (ε) and refractive index (n), which are similar to the above results and confirm the relationship between the band gap and phase transitions of $CsSnI_3$. The obtained results complement the base of scientific works performed in the field of application of perovskite compounds for the development of green energy, and can be used by experimentalists in further studies of crystals and thin films based on $CsSnI_3$.

## 4. Conclusion

Using calculations within the framework of DFT, the issues of structural stability of the $CsSnI_3$ compound and the influence of phase transitions on their electronic and optical properties are considered. Relaxed structures for four phases of $CsSnI_3$ were obtained and their structural stability was assessed from the point of view of comparing the total energy, entropy and enthalpy of their formation. Trends in changes in free energy, entropy, enthalpy, heat of formation energy and heat capacity were justified in terms of the pattern of changes in the total energy of the four phases of $CsSnI_3$ from VASP calculations. The stable phases of $CsSnI_3$ at 0 K are shown and compared. The critical temperatures of phase transitions are found, including the temperature of the black-yellow transformation of $CsSnI_3$. The presence of temperature phase transitions between two orthorhombic phases of $CsSnI_3$ at 360 K was discovered, despite the fact that direct transitions of the α↔γ and γ↔δ types have not yet been reported in any experiment, except for γ→δ transitions under the influence of moisture. Based on the structure of the SCAN obtained after optimization, the bandgap widths for four $CsSnI_3$ phases were calculated and compared with experimental measurements. Shifts in the Fermi level as a result of phase transformations of $CsSnI_3$ are estimated. This study will help to deeply understand the features of the thermodynamic properties of $CsSnI_3$ and list their disadvantages, so that in the future it is advisable to take measures and select alloying elements that stabilize perovskite materials without any gross negative impact on their optoelectronic properties, including the band gap and the ability of good photoabsorption. The results obtained can be useful to experimenters conducting research on the search and creation of materials with desired properties, taking into account the regulation of their band gap and thermodynamic properties.

**Funding**

The work was performed at the S.U. Umarov Physical-Technical Institute of the National Academy of Sciences of Tajikistan with the support of International Science and Technology Center (ISTC), project TJ-2726.